# Engineering electron wavefunctions in asymmetrically confined quasi one-dimensional structures

S. Kumar[1,2,a], M. Pepper[1,2], H. Montagu[1,2], D. Ritchie[3], I. Farrer[3,4], J. Griffiths[3], G. Jones[3]

[1] *Department of Electronic and Electrical Engineering, University College London, Torrington Place, London WC1E 7JE, United Kingdom*

[2] *London Centre for Nanotechnology, 17-19 Gordon Street, London WC1H 0AH, United Kingdom*

[3] *Cavendish Laboratory, J. J. Thomson Avenue, Cambridge CB3 OHE, United Kingdom*

[4] *Department of Electronic and Electrical Engineering, University of Sheffield, Mappin Street, Sheffield S1 3JD, United Kingdom*

We present results on electron transport in quasi-one dimensional (1D) quantum wires in GaAs/AlGaAs heterostructures obtained using an asymmetric confinement potential. The variation of the energy levels of the spatially quantized states is followed from strong confinement through weak confinement to the onset of two-dimensionality. An anticrossing of the initial ground and first excited states is found as the asymmetry of the potential is varied giving rise to two anticrossing events which occur on either side of symmetric confinement. We present results analysing this behaviour and showing how it can be affected by the inhomogeneity in background potential. The use of an enhanced source-drain voltage to alter the energy levels is shown to be a significant validation of the analysis by showing the formation of double rows of electrons which correlate with the anticrossing.

[a] Corresponding author: sanjeev.kumar@ucl.ac.uk

Electron transport in low-dimensional semiconductor nanostructures particularly one-dimensional (1D) systems generates widespread interest among theorists and experimentalists due to the rich fundamental aspects of quantum condensed matter physics. The two-dimensional electron gas (2DEG) formed at the interface of GaAs/AlGaAs heterostructure forms the basis of a typical 1D system.[1-10] Using patterned split gates[3], the 2DEG is electrostatically constricted into a narrow 1D channel, a quantum wire, whose conductance consists of quantised plateaus in units of N.2$e^2$/h, where N=1,2,3… is the number of allowed 1D subbands and the factor 2 is due to spin degeneracy. Generally in a strongly confined 1D system, the simplest approach is to assume a parabolic potential ($V_p(y) = m^*\omega^2 y^2/2$, where $\omega$ is the angular frequency and $m^*$ is the effective mass), with the electrons behaving in a non-interacting manner.[10] In order to search for interaction effects, the system was modified by incorporating a top gate to tailor the carrier concentration within the 1D channel which resulted in many quantum features previously inaccessible[5,7,8,11-14] including the observation of fractionally quantised conductance. In addition to reducing the carrier concentration in the 1D channel, the top gate helps weakening the confinement potential which enhances electron-electron interactions to reconfigure a line of 1D electrons into two rows.[5,7] This behaviour was predicted theoretically on the basis that the increase in the Coulomb interaction ($V_c \sim e^2/r$, where e is the electronic charge and r is the separation between electrons, when $V_c \geq V_p$) would cause electrons to relax sideways[15-17] despite the confinement potential. The net result is a reconfiguration of a line of 1D electrons into a localised zigzag and subsequently into a two-row spatial arrangement, an Incipient Wigner lattice. The zigzag requires electrons in one row to be located midway between the electrons in the other row, an aspect which cannot be confirmed by conductance measurements (supplementary material), however the zigzag scenario of the formation of a new ground state with two centres of charge has been imaged by transverse electron focussing.[18] The zigzag phase would occur when $r_s \cong r_0$, $r_s \sim 1/2n_{1D}$ is the Wigner Seitz radius, $n_{1D}$ is the density of 1D electrons and $r_0 \sim (2e^2/\varepsilon m^*\omega^2)^{3/2}$ is the characteristic length scale, $\varepsilon$ is the dielectric constant, when $V_c \sim V_p$, or 2) $a_B \ll r$, $r_0 \gg 1$, $a_B$ is the Bohr radius.[16,17,19,20] The Wigner state of 1D electrons contains many intriguing spin and charges phases including a suggestion of paired electrons.[13,21,22]

Very recently, a low density, weakly confined, 1D quantum wire displayed fractional conductance plateaus when the confinement potential was tailored to be asymmetric and the system had entered a zigzag phase. This system resulted in a variety of fractional states which were particularly pronounced at 1/2, 1/4, 1/6, 2/5 in the absence of a magnetic field.[8,11]

Numerical calculations performed on modified 1D systems[19,23,24] suggested that the electron density and transverse confinement can be engineered effectively to investigate novel quantum effects.

In this Letter, we show by tuning a variable confinement potential established by asymmetrically biasing a pair of split gates, two gapless states can be produced as a consequence of anticrossing of the initial, non-interacting, ground and first excited states in the 1D constriction.

The experimental devices were fabricated from GaAs/AlGaAs heterostructures grown using molecular beam epitaxy. The layer details for sample 1 and sample 2 are given in supplementary material. The 2DEG formed up to 290 nm beneath the surface had a low-temperature mobility of $1.6 \times 10^6$ cm$^2$/V.s ($2.0 \times 10^6$ cm$^2$/V.s), and an electron density of $9.0 \times 10^{10}$ cm$^{-2}$ ($1.0 \times 10^{11}$ cm$^{-2}$) for sample 1 (sample 2). A pair of split gates of length (width) of 0.4 μm (0.7 μm), and a top gate of length 1 μm separated by a 300 nm thick insulating layer of cross-linked poly(methyl methacrylate) (PMMA) were patterned by a standard lithographic technique on sample 1, whilst the sample 2 was without a top gate. The two-terminal differential conductance ($G$) measurements were performed using an excitation voltage of 10 μV at 73 Hz in a cryofree dilution refrigerator with a lattice temperature of 25 mK.[7,8] In Figs. 2 and 3, the transconductance (d$G$/d$V_{sg}$) of conductance data is plotted in a greyscale plot, where the regions in grey are the risers and black are plateaus in conductance plot, respectively.

Numerical calculations[10,23,25,26] indicate that symmetric biasing the split gates result in a parabolic potential and electrons are strongly confined. However there has not been the same degree of investigation on the effects of decreasing the confinement potential with an asymmetric bias. This is particularly true of the case when the electron gas relaxes and the energy is determined by both the electron-electron repulsion as well as the spatial confinement.[7,8,12]

We establish the asymmetry in the potential by applying a fixed voltage with an offset, $V_A+\Delta V_{SG}$, to one of the split gates, $A$, and the other gate, $B$ whose voltage is $V_B$, is then swept. The offset on $A$ is then further incremented by a fixed value, $\Delta V_{SG}$, and $B$ is then swept as before [Fig 1(a)]. Through this method we are able to create an asymmetric 1D channel by applying an incremental offset voltage on the gate $A$ (gate $B$) and sweeping gate $B$ (gate $A$). This approach will produce a variable confinement potential from being highly asymmetric to symmetric to again highly asymmetric (see supplementary material). Widening the system by a highly asymmetric potential provides a weaker confinement of the electrons enhancing the role of their mutual repulsion compared to the spatial confinement energy. The density distribution in the ground state for a wide channel screens and flattens the potential to produce a dual charge distribution as shown in Fig. 1(c).[16,19,23] Numerical calculations performed in a square well with two particles showed the density distribution dividing into two as a function of the width of the well.[27]

Figure 2 shows conductance characteristics of a 1D quantum wire (sample 1) formed by split gates, the top gate is held at zero voltage. A fixed voltage, $V_A$, with an additional variable offset voltage, $\Delta V_{SG}$, such that $V_A + \Delta V_{SG}$ = -0.4 V (here, $\Delta V_{SG}$=0 V) is applied to gate $A$, and the voltage $V_B$ on gate $B$ is swept to produce the first trace on the left in Fig. 2(a). Next, $\Delta V_{SG}$ is successively incremented negatively by -0.1 V and $B$ is swept to produce successive conductance traces from left to right until the voltage on gate $A$ becomes -5.8 V. Starting from the left, weakly resolved conductance plateaus are seen due to the channel being very wide and close to the 2D limit. At this stage, the system is highly asymmetric with $V_A$=-0.4 V and $V_B$=-5.8 V. Successive traces, formed by relaxing the asymmetry in confinement potential, show the emergence of well-defined plateaus along with a short lived $0.7(2e^2/h)$ feature indicating the system is tending to enter into a comparatively confined regime. Further relaxation of the asymmetry results in plateaus weakening again (traces between $V_B$ of -5.5 to -4.0 V), with the second plateau at $4e^2/h$ weakening faster and before the plateau at $2e^2/h$, particularly shown by the trace at $V_B \sim$ - 4.0 V. On further relaxing the asymmetry, the plateau at $2e^2/h$ disappears, resulting in an anticrossing as seen previously.[5,7] We note that 1) the regular plateaus reappear as the offset on the gate is further increased until the far right where there is a weakening, 2) there is a gradual reduction in spacing between the consecutive conductance traces. This indicates that the electrons are more responsive to the change in split gates voltage as expected if the wavefunction is more extended across the channel so increasing the coupling to the gates. Figure 2(b) is a greyscale plot showing a "well-defined" anticrossing of ground and first excited states illustrated by an arrow in black at $V_A+\Delta V_{SG}, V_B$ (-1.5 V, - 3.0 V). A second anticrossing may be present at $V_A+\Delta V_{SG}, V_B$ (-5.7 V, -0.5 V) but the confinement is now such that levels are very close together approaching two dimensionality. We did not use the top gate in this measurement to avoid any possible electrostatic complication and we establish that the top gate is not essential for the observation of anticrossing, a robust effect seen in top gated, split gates devices.[5-7] There have been previous reports of observations of crossing and anticrossing of energy levels in trenched quantum wires and bilayer systems.[28,29]

We show results from the 1D devices (sample 2) which were without any physical top gate over the split gates in Fig. 3. A fixed voltage with an additional variable offset voltage, $V_B+\Delta V_{SG}$ initially at -1.15 V, corresponding to a highly asymmetric confinement, where $\Delta V_{SG}$=-0.05 V is applied to gate *B*, and the voltage $V_A$ is swept which results in the conductance plots shown in Fig. 3(a). On the left, the system is highly asymmetric as $V_B+\Delta V_{SG}$ =-1.15 V and $V_A$= -5.5 V and successive traces on relaxing the potential asymmetry show the emergence of well-defined plateaus indicating the system is entering into a comparatively stronger confinement regime. On further relaxing (weakening), the asymmetry by increasing the negative offset voltage on gate B (traces between $V_B$ of -4.0 to -3.0 V), in which the second plateau at $4e^2$/h weakens and reappears when the plateau at $2e^2$/h had weakened significantly so giving rise to a direct jump in conductance from 0 to $4e^2$/h, as also seen in Fig. 2. Correspondingly, the greyscale plot in Fig. 3(b) shows a clear anticrossing at $V_A$, $V_B+\Delta V_{SG}$ (-3.3 V, -2.0 V), which we label AX1. On further increasing the negative offset voltage on gate *B*, the regular quantised conductance plateaus emerge until $V_A$~ -1.6 V and $V_B$= -4.5 V. At this stage the spacing between the traces is too narrow, therefore, a magnified view of this section of data is shown in the inset in Fig. 3(b). Remarkably, there is another anti-crossing (AX2) appearing at around $V_A$, $V_B+\Delta V_{SG}$ (-1.6 V, -4.5 V) which was completed unexpected. Both the AXs are appearing due to 1) widening of channel as shown in Fig. 1(c), and (2) an associated enhancement in the importance of the electron-electron interaction. The dotted white line illustrates an almost symmetric potential profile as the voltages on the two gates are almost similar along this line. This indicates that under the given voltage values the system starts with a highly asymmetric potential and with gradual lifting of the asymmetry AX1 appears, further lifting the asymmetry past the anticrossing results in an almost symmetric potential to be felt by the electrons when they are along the white dotted-line regime. Beyond this, an increase in offset applied to gate *B* rebuilds the asymmetry in the system and the system generates a further anticross with the appearance of AX2. However it is not clear why a wide ground state should anticross with the previous ground, and now first excited state. An inspection of the greyscale plot indicates that at a more negative voltage there could be a transition between the first and second excited states. This previous second excited state which is very broad then anticrosses the ground state at AX1 so accounting for the further reduced separation of the plots, such behaviour of excited states has been found in the presence of a strong magnetic field.[6,7] The conductance plateaus in this regime of behaviour weaken at AX2 but do not disappear as they do so near AX1. It is also interesting that the new ground state following the anticross AX1 persists when the potential is symmetric, in agreement with a previous finding that an anticross can occur in a symmetric potential.[5,7] These two effects occurring on either side of symmetric point (dotted white line) indicate the system is very clean, and free from major defects.

To confirm that the effect is not arising from an impurity potential we repeated the measurement by swapping the gates, i.e, now gate *B* is sweeping and the offset is applied on gate *A*. Similar results were obtained with two sets of anticrossings AX3 and AX4 (see supplementary material).

Figure 4 shows the effect of applying a fixed dc source-drain bias of 3 mV on the conductance characteristics in sample 2, where gate *A* is swept whilst an incremental offset is applied on gate *B*. Previous work has shown that measurements of differential conductance in the presence of an enhanced source-drain voltage has revealed that both spin and momentum degeneracy are lifted.[30] In Fig. 4, the first trace on the left is where the offset on gate *B* is set at $V_B+\Delta V_{SG} = -1.5$ V. The $0.25(2e^2/h)$ plateau persists despite increasing offset on the gate *B* or relaxing the asymmetric potential until around $V_A \sim -3.5$ V when an additional plateau at $0.5(2e^2/h)$ was observed (blue trace). This value of gate voltage is where the anticross occurs in the ohmic regime with the weak first plateau being observed at $2e^2/h$ and the $0.5(2e^2/h)$ absent, this further confirms that the confinement potential at the anticross results in two rows of charge each having a conductance of $0.25(2e^2/2h)$ as seen in Refs. [6,7]. The existence of the two separate rows indicates that they do not couple and their subsequent disappearance beyond the region of anticross implies that the two centres of charge have now linked in one wavefunction, i.e. the rows have become coupled.

The interaction between electrons and subsequent entanglement in closely separated rows such as ladder compounds is a topic of considerable interest. It has been suggested that the interaction between two rows in the quasi-1D system as described here is responsible for the observation of non-magnetic fractional quantisation of conductance.[8,11,12] The manner in which rows form and their subsequent behaviour, such as whether completely separate or coupled in a single wavefunction, is therefore of considerable importance in understanding the mechanism underlying fractional quantisation. In addition, fractional plateaus have been observed[12] which possess even denominators pointing to the existence of electron pairing as a result of the row formation. Such pairing appears a result of entanglement indicated by formation of a single wavefunction involving two centres of charge. It has been shown previously that a zigzag phase can be traced and imaged using transverse electron focusing where subpeaks corresponding to two centres of electron density are observed in the first focusing peak.[18] This result combined with the analysis of conductance shown here, and observation of row formation, are very strong evidence for the confirmation of theoretical suggestions of rich spin and charge phases in the zigzag.[13,14,16,17]

In conclusion, we have investigated the engineering of electron wavefunctions in quasi-1D channels. We have shown that there are two sets of anticrossings on either side of a symmetric, wide 1D channel as a function of manipulating asymmetry in the confinement potential. The formation of separate rows is confirmed by measurements of the non-ohmic differential conductance. We suggest exciting 1D quantum physics can be exploited using confinement manipulation which enables us to follow the energy levels from the single electron situation where the levels are determined by the size quantisation through the increasingly interaction dominated regime. The form of the wavefunctions and their energetics then deviate strongly from the independent electron behaviour with a drastic change of ground state and modification of the ground wavefunction which becomes more laterally extended in the strongly interacting regime.

**Data Availability**

The data that support the findings of this study are available from the corresponding author upon reasonable request.

See supplementary material (below) for additional data to support experimental results.

**Acknowledgements**

Authors are grateful to the late Prof K Berggren and Prof S Bose for many fruitful discussions. The work is funded by the Engineering and Physical Sciences Research Council (EPSRC), United Kingdom Research and Innovation (UKRI), UK and UKRI Future Leaders Fellowship (Reference: MR/S015728/1).

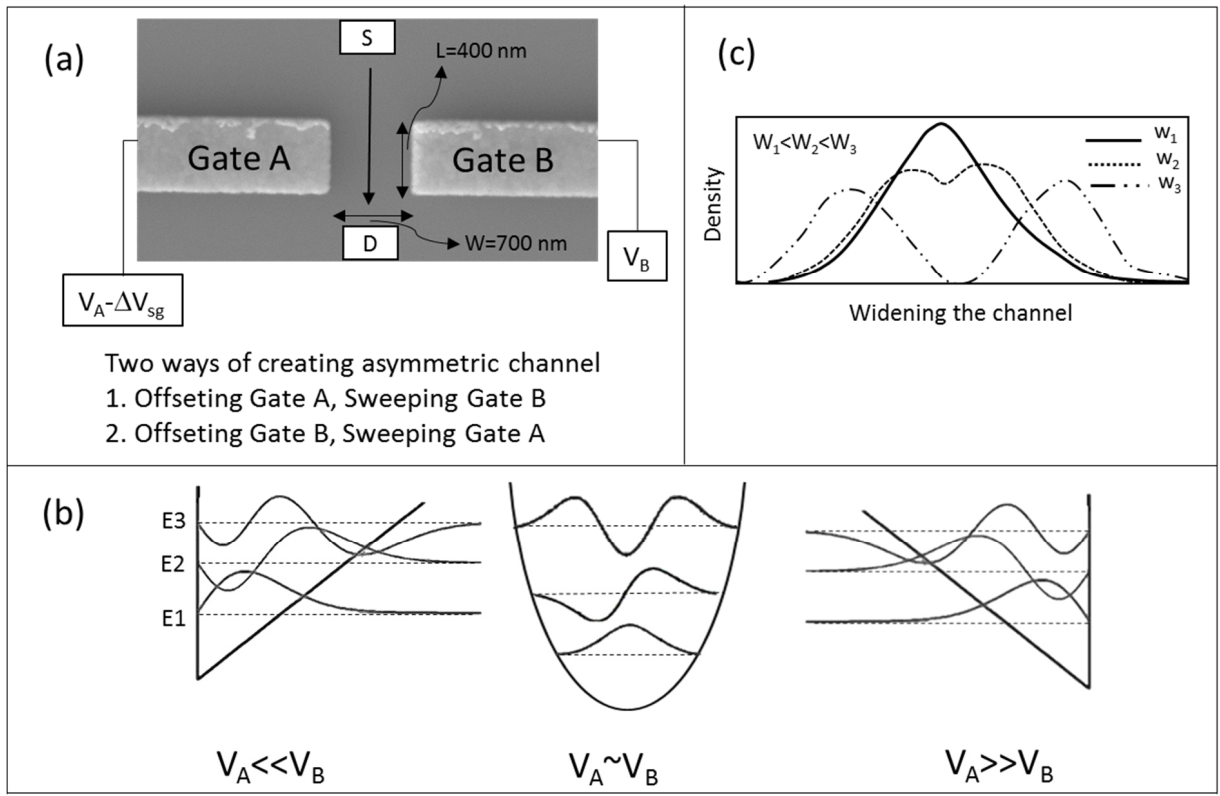

Figure 1: (a) A schematic of asymmetric biasing on a pair of gates *A* and *B*. The transport direction is from source, *S* towards drain, *D*. The length, *L* of 1D channel is 400 nm and the width, W is 700 nm. (b) An artistic approach of a possible variation of potential profile taking place in the system. The left triangular shape profile shows a possible case when the gate *A* is acting like a hard wall and the gate B produces a varying potential. By offsetting gate *A* further a symmetric parabolic potential can be achieved when equal confinement voltages are applied. The reverse situation of the triangular profile as seen on the right is achieved where gate *B* is now acting as a hard wall. (c) A cartoon to explain a possible mechanism where the density distribution takes different form in a flat bottom potential as a function of increasing width of channel. Here, $W_1 < W_2 < W_3$.

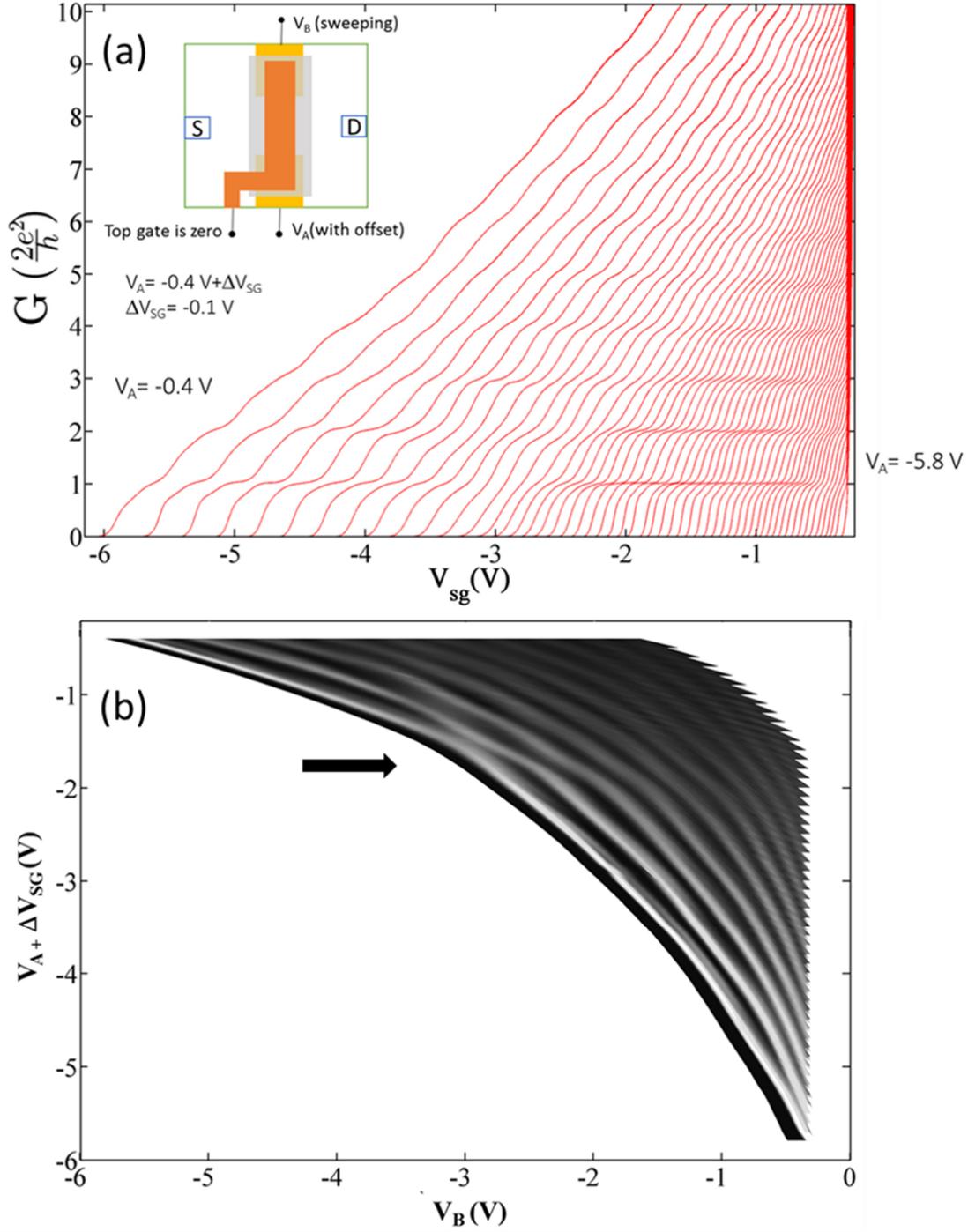

Figure 2: (a) Conductance characteristics measured by sweeping voltage on gate *B*, $V_B$ as a function of increasing negative offset voltage ($\Delta V_{SG}$=-0.1 V) on gate *A* so that $V_A$ ranges from -0.4 V(left) to -5.8 V(right). Inset shows a device schematic where split gates are shown in yellow and a top gate over the split gates in orange. (b) A greyscale plot showing an anticrossing of the ground and first excited states at $V_A+\Delta V_{SG}, V_B$ (-1.5 V, -3.0 V) as indicated by a black arrow.

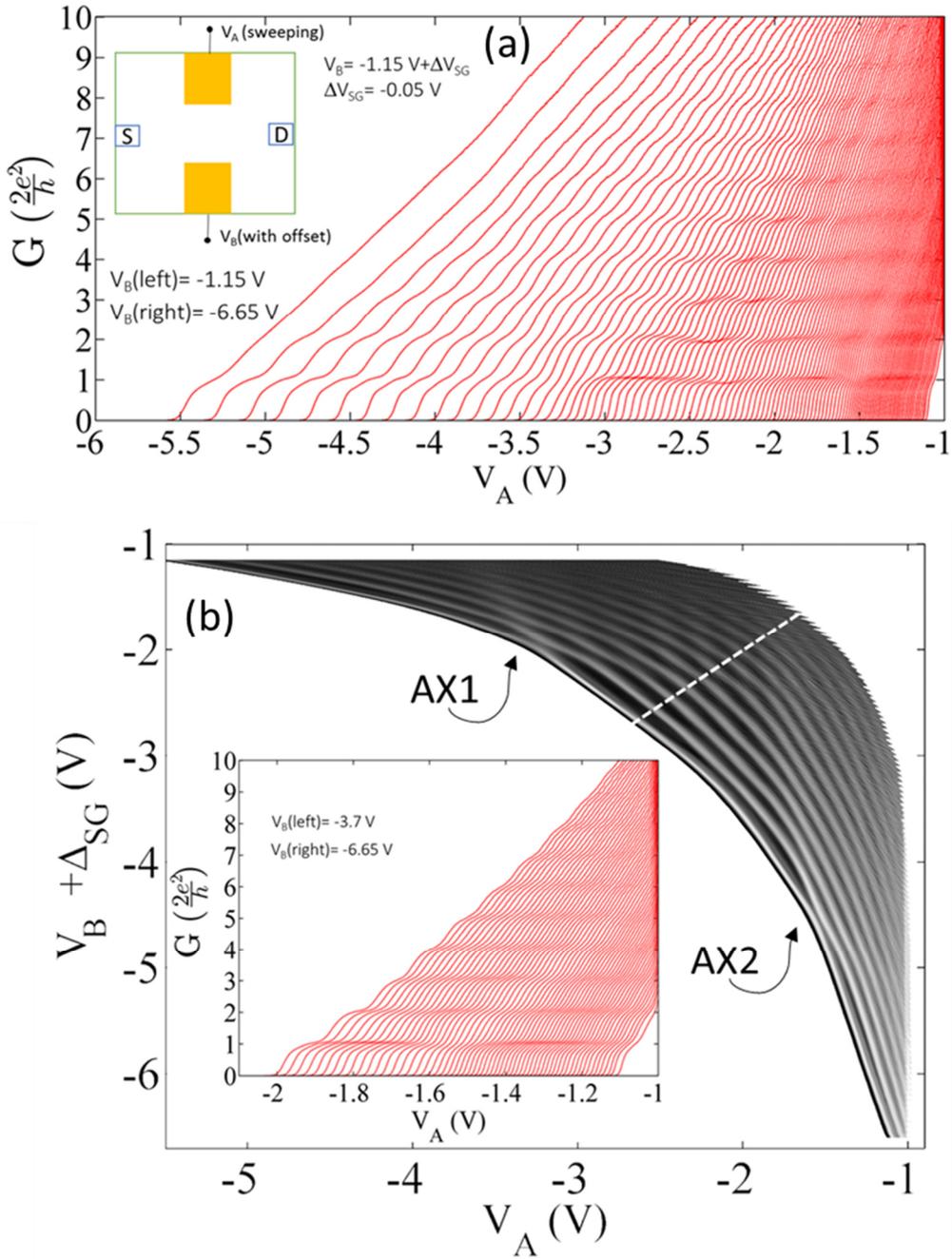

Figure 3:(a) Conductance characteristics measured by sweeping voltage on gate $A$, $V_A$ as a function of increasing negative offset voltage ($\Delta V_{SG}$= -0.05 V) on gate $B$ so that $V_B$ ranges from -1.15 V(left) to – 6.65 V(right). Inset shows device schematic where split gates are shown in yellow. (b). A greyscale plot showing two sets of anticrossing of the ground and first excited states, one at $V_A$, $V_B+\Delta V_{SG}$ (-3.3 V, -2.0V) and the other at $V_A$, $V_B+\Delta V_{SG}$ (-1.6 V, -4.5 V). A dotted white line through the greyscale plot shows the subbands which are experiencing a symmetric potential. Inset shows a magnified view of conductance data shown in (a) where $V_A$ ranges from ~ -2.1 to -1.0 V.

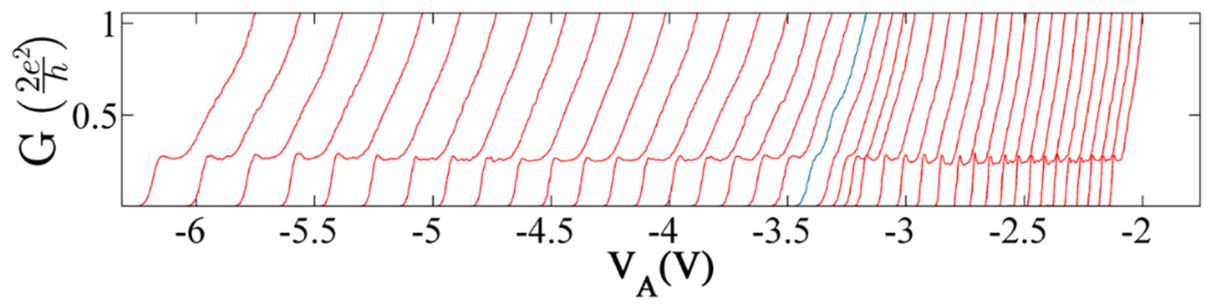

Figure 4: The effect of a fixed source drain bias, $V_{sd}$= 3 mV on the differential conductance of a device from sample 2 by sweeping gate *A* and offsetting gate *B (i.e. $V_B+\Delta V_{SG}$)* from -1.5 V(left) to -4.4 V (right) with increments of $\Delta V_{SG}$=-0.06 V.

# Engineering electron wavefunctions in asymmetrically confined quasi one-dimensional structures

S. Kumar[1,2], M. Pepper[1,2], H. Montagu[1,2], D. Ritchie[3], I. Farrer[3,4], J. Griffiths[3], G. Jones[3]

[1] *Department of Electronic and Electrical Engineering, University College London, Torrington Place, London WC1E 7JE, United Kingdom*

[2] *London Centre for Nanotechnology, 17-19 Gordon Street, London WC1H 0AH, United Kingdom*

[3] *Cavendish Laboratory, J. J. Thomson Avenue, Cambridge CB3 0HE, United Kingdom*

[4] *Department of Electronic and Electrical Engineering, University of Sheffield, Mappin Street, Sheffield S1 3JD, United Kingdom*

## Experimental

The devices used in the present work were fabricated from GaAs/AlGaAs heterostructures grown using molecular beam epitaxy. The layer structure for sample 1 and sample 2 starting from the top, cap layer is as follows: Sample 1: GaAs cap(10 nm)/Al$_{0.33}$Ga$_{0.67}$As (doping 8x10$^{16}$ cm$^{-2}$) (200 nm)/Al$_{0.33}$Ga$_{0.67}$As (75 nm)/GaAs (buffer)(1000 nm). Sample 2: GaAs cap(10nm)/Al$_{0.33}$Ga$_{0.67}$As(200nm)/GaAs(0.56nm)/As:Si(8x10$^{11}$cm$^2$)/GaAs(0.56nm) /Al$_{0.33}$Ga$_{0.67}$As(75nm)/GaAs(buffer) (1000nm). The 2DEG formed up to 290 nm beneath the surface had a low-temperature mobility of 1.6 × 10$^6$ cm$^2$/V.s (2.0 × 10$^6$ cm$^2$/V.s), and an electron density of 9.0 × 10$^{10}$ cm$^{-2}$ (1.0 × 10$^{11}$ cm$^{-2}$) for sample 1 (sample 2). A pair of split gates of length (width) of 0.4 μm (0.7 μm), and a top gate of length 1 μm separated by a 300 nm thick insulating layer of cross-linked poly(methyl methacrylate) (PMMA) were patterned by a standard lithographic technique on sample 1, whilst the sample 2 was without a top gate.

## Asymmetric Potential

We establish the asymmetry in the potential by applying a fixed offset, $V_A$+$\Delta V_{SG}$, to one of the split gates, $A$, and the other gate, $B$ whose voltage is $V_B$, is then swept until the channel is pinched off. The offset on $A$ is then further incremented by a fixed value, $\Delta V_{SG}$, and $B$ is then swept as before. This is represented diagrammatically in Fig 1(a) in the main text. Through this method we are able to create an asymmetric 1D channel by applying an incremental offset voltage on the gate A (gate B) and sweeping gate B (gate A). This approach is expected to produce an asymmetric confinement potential which cannot be produced when both gates are symmetrically biased. In the limit the modified 1D system may be pictured as a modified triangular or sawtooth potential comprising hard wall on one side with the other side more gradual. Initially, gate A is fixed at a voltage less than the definition voltage where the 1D-2D transition takes places. Gate B is then swept until the channel is pinched off, this process will require a comparatively larger negative voltage to pinch off the channel as gate A is held at a less negative gate voltage. As the negative offset voltage is incremented a situation can arise when the voltage on both the gates would be almost similar producing the symmetric parabolic potential. Continuing the addition of a negative offset to A will make it negative compared to gate B, a reverse situation to the initial. At this stage, the potential would again become highly asymmetric however now gate A will form the hard wall part of the potential and gate B the slowly varying part. This is represented in Fig. 1(b) through a cartoon in the main text. This process will continue until the system transitions though a slowly changing confinement potential (from being highly asymmetric to symmetric to again highly asymmetric). Widening the system by a highly asymmetric potential provides a weaker confinement to the electrons enhancing the role of their mutual repulsion compared to the spatial confinement energy.

## Conductance in an Asymmetric Potential

Figure S1(a) shows the results obtained when role of gates is reversed in comparison to Fig. 3. Similar results are obtained as before and two sets of anticrossings are reproduced, as expected, AX3 ($V_A$, $V_B$: -1.52 V, -4.40 V) and AX4 ($V_A$, $V_B$: -3.20 V, -1.87 V) as shown in the transconductance plot in Fig. 1S(b). The dotted white line shows the points of symmetric potential. The voltage values of AXs in two results in Fig. 3 and Fig. S1 are different which may be expected in mesoscopic systems where there may be an inhomogeneity in the background potential, however clearly this does not affect the phenomena being reproduced.

We note that the AX3 is weaker than its counterpart in Fig. 3 although AX4 seems extremely strong involving 3 subbands. Following AX3 the individual plots are located even closer together indicating a wavefunction evenly distributed although perhaps the hybridisation has resulted in charge enhancement close to the split gates.

## Double rows and the Zigzag

The formation of a zigzag does not necessarily lead to a change in conductance as in the first stage it is a distorted line with a conductance of $2e^2/h$. Further decrease, (increase), in confinement, (repulsion), can lead to a splitting into two independent rows with the first plateau appearing at $4e^2/h$, which corresponds to a zigzag with anti-bonding orbitals. It is interesting to note in this context that a strong magnetic field which lifts the spin degeneracy produces a crossing of the initial ground and first excited states and a conductance of $2e^2/h$, as expected for two separate spin polarised lines. Following the crossing the new ground state has a conductance of $e^2/h$, corresponding to a single wavefunction although the dependence on split gate voltage indicates that the wavefunction is now more extended across the channel – as expected. As seen in the greyscale plots the former first excited state drops below the former ground state to become the new ground state and presumably the hybridisation removes the node in the wavefunction so lowering the energy. Although the unperturbed ground and first excited wavefunctions are orthogonal they will be so distorted by the strong interaction that hybridisation can occur so producing the observed anticrossing.

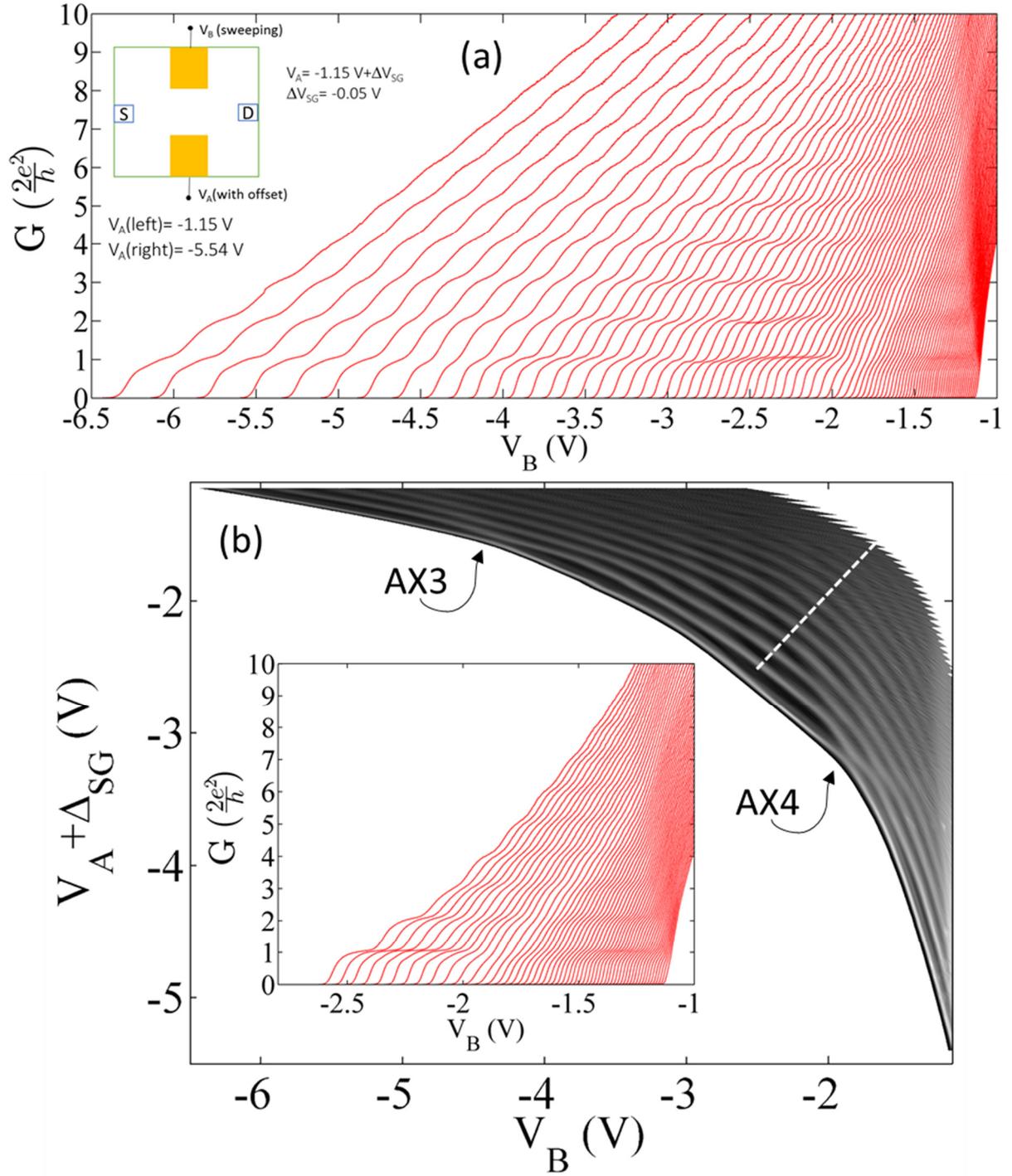

Figure S1:(a) Conductance plot of a 1D quantum wire from sample 2 measured by sweeping voltage on gate B, $V_B$ as a function of increase in offset voltage ($\Delta V_{SG}$=-0.05 V) on gate A so that the voltage on gate A, $V_A$ ranges from -1.15 V on the left to – 5.54 V on the right. The inset shows device schematic where split gates are shown in yellow. S and D represent the source and drain. (b) A greyscale plot of transconductance of data shown in (a), here the regions in grey are the risers and back are plateaus in conductance plot, respectively. Two sets of anticrossing of the ground state and first excited state regimes are observed, one at $V_A, V_B$ (-1.52 V, -4.4 V) and the other at $V_A, V_B$ (-3.2 V, -1.87 V). A dotted white line through the greyscale plot shows the subbands which are experiencing symmetric potential

(i.e, both $V_A$ and $V_B$ are approximately similar values). Inset shows a magnified view of conductance data shown in (a) where $V_B$ ranges from ~ -2.8 to -1.0 V.